# Ensemble Dependence of the Transient Fluctuation Theorem

Debra J. Searles
Department of Chemistry
University of Queensland
Brisbane, Qld 4072 Australia

Denis J. Evans
Research School of Chemistry
Australian National University
Canberra, ACT 0200 Australia

## Abstract

The Fluctuation Theorem gives an analytical expression for the probability of observing second law violating dynamical fluctuations, in nonequilibrium systems. At equilibrium statistical mechanical fluctuations are known to be ensemble dependent. In this paper we generalise the Transient and Steady State Fluctuation Theorems to various nonequilibrium dynamical ensembles. The Transient and Steady State Fluctuation Theorem for an isokinetic ensemble of isokinetic trajectories is tested using nonequilibrium molecular dynamics simulations of shear flow.

## 1. INTRODUCTION

The fluctuation theorem[1-3] (FT) gives a general formula for the logarithm of the probability ratio that the time averaged dissipative flux takes on a value, $\bar{J}_t$, to minus the value, namely $-\bar{J}_t$, in a nonequilibrium steady state. This formula is an analytic expression that gives the probability, for a finite system and for a finite time, that the dissipative flux flows in the reverse direction to that required by the Second Law of Thermodynamics. For steady state trajectories, the Steady State FT (SSFT) is only true in the long time limit. Evans and Searles[2] have shown that if transient trajectories are considered rather than steady state trajectories, a Transient FT (TFT) that is true at all times can be derived. Further, when the nonequilibrium steady state is unique, one would expect the asymptotic convergence of the Transient to the Steady State Fluctuation Theorem since averages over transient segments should then approach those taken over nonequilibrium steady state segments. However, there has been some recent discussion[4,5] of this point and not all parties agree on asymptotic convergence.

The Transient FT considers the thermostatted response of an ensemble of systems to an applied dissipative field. The system is thermostatted so that it may reach a steady state after a Maxwell time, $\tau_M$. In the original derivation[2], it was supposed that the initial ensemble was the microcanonical ensemble and the dynamics was isoenergetic. This initial ensemble was chosen because the probability of observing trajectories originating in a specified phase volume is simply proportional to the measure of that volume. However, it is straightforward to apply the same procedure to an arbitrary initial ensemble, and an analytical form of the FT that is valid at all times can be obtained. In this note we demonstrate how the FT can be extended to an arbitrary system and as an example, derive a transient FT for the isokinetic nonequilibrium response of an initial isokinetic ensemble.

## 2. TRANSIENT FLUCTUATION THEOREM (TFT)

Consider an N-particle system with coordinates and peculiar momenta, $\{\mathbf{q}_1,\mathbf{q}_2,..\mathbf{q}_N,\mathbf{p}_1,..\mathbf{p}_N\} \equiv (\mathbf{q},\mathbf{p}) \equiv \boldsymbol{\Gamma}$. The internal energy of the system is $H_0 \equiv \sum_{i=1}^{N} p_i^2/2m + \Phi(\mathbf{q}) = K + \Phi$ where $\Phi(\mathbf{q})$ is the interparticle potential energy, which is a function of the coordinates of all the particles, $\mathbf{q}$ and K is the total peculiar kinetic energy. In the presence of an external field $F_e$, the thermostatted equations of motion are taken to be,

$$\begin{aligned} \dot{\mathbf{q}}_i &= \mathbf{p}_i / m + \mathbf{C}_i(\boldsymbol{\Gamma})F_e \\ \dot{\mathbf{p}}_i &= \mathbf{F}_i(\mathbf{q}) + \mathbf{D}_i(\boldsymbol{\Gamma})F_e - \alpha(\boldsymbol{\Gamma})\mathbf{p}_i \end{aligned} \tag{1}$$

where $\mathbf{F}_i(\mathbf{q}) = -\partial\Phi(\mathbf{q}) / \partial\mathbf{q}_i$ and $\alpha$ is the thermostat multiplier which in this case is applied to the peculiar momenta, and $\mathbf{C}_i$ and $\mathbf{D}_i$ represent the coupling of the system to the field. The dissipative flux is given by J where $\dot{H}_0^{ad} \equiv -J(\boldsymbol{\Gamma})VF_e$.

The probability that a trajectory segment will be observed within an infinitesimal phase space volume of size $\delta V$ about $\boldsymbol{\Gamma}$ at time t, $\Pr(\delta V(\boldsymbol{\Gamma}(t),t))$ is given by,

$$\Pr(\delta V(\boldsymbol{\Gamma}(t),t)) = f(\boldsymbol{\Gamma}(t),t)\delta V(\boldsymbol{\Gamma}(t),t). \tag{2}$$

where $f(\boldsymbol{\Gamma}(t),t)$ is the normalised phase space distribution function at the point $\boldsymbol{\Gamma}(t)$ at time t. The Lagrangian form of the non-equilibrium Kawasaki distribution function[2] is given by:

$$f(\boldsymbol{\Gamma}(t),t) = \exp\left[-\int_0^t \Lambda(\boldsymbol{\Gamma}(s))ds\right] f(\boldsymbol{\Gamma}(0),0) = \exp\left[-\int_0^t \Lambda(\boldsymbol{\Gamma}(s))ds\right] f(\boldsymbol{\Gamma}(0),0) \tag{3}$$

where $\Lambda(\boldsymbol{\Gamma}) \equiv \partial\dot{\boldsymbol{\Gamma}} \bullet / \partial\boldsymbol{\Gamma}$ is the phase-space compression factor. Now consider the set of initial phases in the volume element of size $\delta V(\boldsymbol{\Gamma}(0),0)$ about $\boldsymbol{\Gamma}(0)$. At time t, these phases will occupy a volume $\delta V(\boldsymbol{\Gamma}(t),t)$. Since the number of ensemble members within a comoving phase volume is conserved, (3) can be used to show,

$$\delta V(\boldsymbol{\Gamma}(t),t) = \exp\left[\int_0^t \Lambda(\boldsymbol{\Gamma}(s))ds\right] \delta V(\boldsymbol{\Gamma}(0),0). \tag{4}$$

which is simply the phase space volume contraction along the trajectory, from $\boldsymbol{\Gamma}(0)$ to $\boldsymbol{\Gamma}(t)$.

We will refer the to the trajectory starting at $\boldsymbol{\Gamma}(0)$ and ending at $\boldsymbol{\Gamma}(t)$ as $\boldsymbol{\Gamma}_{(0;t)}$. As discussed previously[2], a time-reversed trajectory segment that is initiated at time zero, $\boldsymbol{\Gamma}^*_{(0;t)}=M^{(T)}(\boldsymbol{\Gamma}_{(0;t)})$ where $M^{(T)}$ is the time-reversal mapping, can be constructed by applying a time-reversal mapping at the midpoint of this trajectory and propagating forward and backward in time from this point for a period of t/2 in each direction. See [2] for further details. The point $\mathbf{\Gamma}^*(0)$ is related to the point $\mathbf{\Gamma}(t)$ by a time-reversal mapping. Since the Jacobian of the time-reversal mapping is unity, the phase volume $\delta V(\mathbf{\Gamma}(t),t)$ is equal to the phase volume $\delta V(\mathbf{\Gamma}^*(0),0)$. The ratio of the probability of observing the two volume elements at time zero is:

$$\begin{aligned}\frac{\Pr(\delta V(\mathbf{\Gamma}(0),0))}{\Pr(\delta V(\mathbf{\Gamma}^*(0),0))} &= \frac{f(\mathbf{\Gamma}(0),0)\delta V(\mathbf{\Gamma}(0),0)}{f(\mathbf{\Gamma}^*(0),0)\delta V(\mathbf{\Gamma}^*(0),0)} \\ &= \frac{f(\mathbf{\Gamma}(0),0)}{f(\mathbf{\Gamma}(t),0)}\exp\left[-\int_0^t \Lambda(\mathbf{\Gamma}(s))ds\right]\end{aligned} \tag{5}$$

where we have used the symmetry of the time-reversal mapping and equation (4) to obtain the final equality. This TFT is completely general and applies to any ensemble or type of dynamics. If the initial phase space distribution function is known (regardless of whether it is an equilibrium distribution), we can then obtain an analytical expression for the probability ratio. Note that the phase space distribution function in the numerator and denominator both refer to that at time zero, therefore it is readily applied to a system which is at equilibrium at time zero but moves away from equilibrium (when the distribution function may be intractable).

For arbitrary initial ensembles and arbitrary dynamics (constant energy, temperature and/or pressure etc) it is convenient to define a general dissipation function $ß(\mathbf{\Gamma})$, so that

$$\begin{aligned}\int_0^t ds\ ß(\mathbf{\Gamma}(s)) &= \ln\left(\frac{f(\mathbf{\Gamma}(0),0)}{f(\mathbf{\Gamma}(t),0)}\right) - \int_0^t \Lambda(\mathbf{\Gamma}(s))ds \\ &\equiv \overline{ß}_t t\end{aligned} , \tag{6}$$

we can obtain a formula for the probability ratio of observing a particular value of $\overline{ß}_t$ and its negative. This is achieved by summing over all appropriate regions of phase and it is straightforward to show that a TFT of the form,

$$\ln \frac{\Pr(\overline{ß}_t = A)}{\Pr(-\overline{ß}_t = A)} = At \qquad (7)$$

for the property ß is obtained.[2,7]

The equation (7) has previously been considered for a system that is initially in a microcanonical ensemble and which undergoes isoenergetic dynamics.[2,6-8] In this case $ß(\boldsymbol{\Gamma}) = -\Lambda(\boldsymbol{\Gamma}) = -VF_e\beta J(\boldsymbol{\Gamma})$ and simulations have verified the resulting TFT, that is $\ln \frac{\Pr(\overline{\Lambda}_t = A)}{\Pr(-\overline{\Lambda}_t = A)} = -At = -VF_e\beta\overline{J}_t t$ where $\overline{J}_t = A/(VF_e\beta)$.[1-2,5,7,8] The purpose of this paper is to derive equations equivalent to (7) for other ensembles and dynamics.

As an example, we consider a system initially in the the isokinetic ensemble and undergoing isokinetic dynamics. The isokinetic distribution function is,

$$f(\boldsymbol{\Gamma}(0),0) = f_K(\boldsymbol{\Gamma}(0),0) = \frac{\exp(-\beta H_0(\boldsymbol{\Gamma}(0)))\delta(K(\boldsymbol{\Gamma}(0)) - K_0)}{\int d\boldsymbol{\Gamma} \exp(-\beta H_0(\boldsymbol{\Gamma}))\delta(K(\boldsymbol{\Gamma}) - K_0)}. \qquad (8)$$

Substituting into equation (5) gives

$$\frac{\Pr(\delta V(\boldsymbol{\Gamma}(0),0))}{\Pr(\delta V(\boldsymbol{\Gamma}^*(0),0))} = \frac{f_K(\boldsymbol{\Gamma}(0),0)\delta V(\boldsymbol{\Gamma}(0),0)}{f_K(\boldsymbol{\Gamma}^*(0),0)\delta V(\boldsymbol{\Gamma}^*(0),0)}$$

$$= \frac{\exp(-\beta H_0(\boldsymbol{\Gamma}(0)))}{\exp(-\beta H_0(\boldsymbol{\Gamma}(t)))} \exp\left[-\int_0^t \Lambda(\boldsymbol{\Gamma}(s))ds\right] \qquad (9)$$

$$= \exp(\beta \int_0^t \dot{\Phi}(\boldsymbol{\Gamma}(s))ds) \exp\left[-\int_0^t \Lambda(\boldsymbol{\Gamma}(s))ds\right]$$

where we have used the symmetry of the mapping, $H_0(\boldsymbol{\Gamma}^*(0)) = H_0(\boldsymbol{\Gamma}(t))$, $V(\boldsymbol{\Gamma}^*(0),0) = V(\boldsymbol{\Gamma}(t),t)$ and $K(\boldsymbol{\Gamma}^*(0),0) = K(\boldsymbol{\Gamma}(t),t)$ to obtain the second equality and that $H_0(\boldsymbol{\Gamma}(t)) = H_0(\boldsymbol{\Gamma}(0)) + \int_0^t \dot{H}_0(\boldsymbol{\Gamma}(s))ds = H_0(\boldsymbol{\Gamma}(0)) + \int_0^t \dot{\Phi}(\boldsymbol{\Gamma}(s))ds$ to obtain the final equality. We see

that:

$$ß(\mathbf{\Gamma}) = \beta\dot{\Phi}(\mathbf{\Gamma}) - \Lambda(\mathbf{\Gamma})$$

$$= -\beta J(\mathbf{\Gamma})VF_e$$

and from (7) we therefore have,

$$\ln\frac{\Pr(\bar{J}_t = A)}{\Pr(\bar{J}_t = -A)} = -AtF_e\beta V. \tag{10}$$

The TFT given by equation (10) is true at all times for the isokinetic ensemble when all initial phases are sampled from an equilibrium isokinetic ensemble.

If the system reaches a unique steady state, then at long times the value of J will fluctuate about its steady state value. A set of nonequilibrium time averaged currents, $\{\bar{J}_{t,SS}\}$, can be generated by evolving time along a single phase space trajectory which starts at some initial phase which is consistent with the macroscopic conditions (N,V,T or E, Fe, etc). One waits for a time which is much longer than the Maxwell time, $\tau_M$, which characterises the transient response, before one begins to analyse time contiguous "steady state" trajectory segments and computes for example the statistics of the set $\{\bar{J}_{t,SS}\}$.

If the steady state is unique, the statistics of the set $\{\bar{J}_{t,SS}\}$ is independent of the original initial phases. Since the steady state is unique we can also gather the statistical information on $\{\bar{J}_{t,SS}\}$ by studying the response of an ensemble of initial phases, provided we wait several Maxwell times before gathering "steady state" data. After the transient response time, the instantaneous system properties will be characteristic of the steady state. The set $\{\bar{J}_{t,T}\}$ averaged over transient trajectory segments, starting from an *equilibrium* ensemble of phases will approach the (unique) *steady state* set $\{\bar{J}_{t,SS}\}$ with $\{\bar{J}_{t,T}\} = \{\bar{J}_{t,SS}\} + O(\tau_M / t)$.

The transient flux, $\bar{J}_{t,T} = \bar{J}_{t,SS} + O(\tau_M / t)$, can be expanded using a Taylor series analysis and because at sufficiently long times the deviation of $\bar{J}_{t,SS}$ from its mean value decreases as $O(t^{-1/2})$, equation (10) can be written,

$$\ln\frac{\Pr(\bar{J}_{t,T}=A)}{\Pr(\bar{J}_{t,T}=-A)}=\ln\frac{\Pr(\bar{J}_{t,SS}=A)}{\Pr(\bar{J}_{t,SS}=-A)}+\ln\left(O\left(\tau_M/t\right)t\right)=-AtF_e\beta V. \qquad (11)$$

Therefore since $t_M$ is finite, provided t is sufficiently long and the steady state is unique we expect,

$$\lim_{t\to\infty}\frac{1}{t}\ln\frac{\Pr(\bar{J}_{t,SS}=A)}{\Pr(\bar{J}_{t,SS}=-A)}=-AF_e\beta V. \qquad (12)$$

Although there is some contention regarding this point[4,5], the numerical tests presented below support the possibility of this convergence.

In Table I we give Transient FTs for various ergodically consistent ensembles[6], that is for systems where the zero field dynamics preserves the initial ensemble: $\partial f(\mathbf{\Gamma},0)/\partial t\,|_{F_e=0}=0$. We also give the exact form of the steady state FT derived from the transient FT by assuming the steady state is unique.

The steady state FT, proposed and tested by Evans, Cohen and Morriss[1] and proven by Gallavotti and Cohen referred to particular conditions (that the dynamics is dissipative, isoenergetic,[9] reversible and chaotic) and may be expressed by the formula[4]

$$\lim_{t\to\infty}\frac{1}{t}\ln\frac{\Pr(\bar{\Lambda}_t=A)}{\Pr(\bar{\Lambda}_t=-A)}=-A. \qquad (13)$$

In a Gaussian isoenergetic system the phase space contraction rate is instantaneously proportional to both the entropy production rate per unit volume and the dissipative flux, and therefore the fluctuations in these three properties will be directly related. Alternative forms of the FT for the isoenergetic system in terms of the fluctuations of the dissipative flux or the entropy production rate per unit volume are therefore trivially obtained. In the original work of Gallavotti and Cohen, for example, the FT was expressed in terms of the entropy production rate per unit volume.[3] For a system undergoing *isokinetic* dynamics the dissipative flux is no longer instantaneously proportional to the phase space contraction rate (although the time averaged values are proportional in the limit

t→∞ since $\overline{\Lambda}_t(\mathbf{\Gamma}) = VF_e\beta\bar{J}_t(\mathbf{\Gamma}) + O(1/t))$, and it is therefore of interest to consider (13) for the isokinetic dynamics.

## 3. NUMERICAL RESULTS FOR TRANSIENT AND STEADY STATE SYSTEMS

We test equations (10) and (13) for systems consisting of N=32 WCA particles in two Cartesian dimensions undergoing isokinetic shear flow in two Cartesian dimensions using the SLLOD equations of motion for planar Couette flow.[6] The equations of motion are:

$$\begin{aligned} \dot{\mathbf{q}}_i &= \mathbf{p}_i/m_i + \mathbf{i}\gamma y_i \\ \dot{\mathbf{p}}_i &= \mathbf{F}_i - \mathbf{i}\gamma p_{yi} - \alpha\mathbf{p}_i \end{aligned} \tag{14}$$

where $\gamma$ is the applied strain rate, and $\alpha$ is the Gaussian isokinetic thermostat multiplier:

$$\alpha = \frac{\sum_{i=1}^{N} \mathbf{F}_i \cdot \mathbf{p}_i - \gamma p_{xi} p_{yi}}{\sum_{i=1}^{N} \mathbf{p}_i \cdot \mathbf{p}_i}. \tag{15}$$

In this case the dissipative flux, J is equal to $P_{xy}$, the xy element of the pressure tensor and $\Lambda(\mathbf{\Gamma}) = -2N\alpha(\mathbf{\Gamma}) + O(1)$. All results below are presented in reduced units. The simulations are carried out for systems under two conditions. In the first case a temperature of T = 1.0, a particle density of n = 0.8 and an applied strain rate of $F_e = \gamma = 0.5$ are employed. This applied strain rate is sufficiently high that we are in the nonlinear regime and the shear thinning is approximately 12%. In the second case a temperature of T = 1.0, a particle density of n = 0.4 and an applied strain rate of $F_e = \gamma = 0.01$ are employed and the system is in the linear regime.

For both sets of conditions, transient trajectories and steady state trajectory segments are considered. The transient trajectories of length t are initiated from an isokinetic equilibrium ensemble and the time-averaged value of the thermostat multiplier and the dissipative flux is calculated for each trajectory. The steady state trajectory segments are initiated at equally spaced time origins along a single steady state trajectory. In this case the thermostat multiplier and the dissipative flux are calculated for each trajectory segment of length t.

Results for transient trajectories are shown in Figures 1 and 2. In Figure 1, transient trajectory segments are of length t = 0.6 are considered. The value of $W_J(t) \equiv -\ln\left(\Pr(\bar{J}_t)/\Pr(-\bar{J}_t)\right)/(t\gamma\beta V)$ is plotted as a function of $\bar{J}_t$ to test equation (10), and $W_\alpha(t) \equiv \ln\left(\Pr(\bar{\alpha}_t)/\Pr(-\bar{\alpha}_t)\right)/(2Nt)$ is plotted as a function of $\bar{\alpha}_t$ to test equation (13). If the equations are valid, a straight line of unit slope is obtained. As expected, the results indicate that equation (10) is valid whereas equation (13) is clearly **not** valid for an averaging time t = 0.6. In Figure 2 we show how the slopes of the lines formed from a plot of $W_J(t)$ versus $\bar{J}_t$ and $W_\alpha(t)$ versus $\bar{\alpha}_t$ vary with averaging time, t. We denote the slopes of these lines $S_J(t)$ and $S_\alpha(t)$ respectively. This figure demonstrates that the Transient FT derived in this paper for an isokinetic systems and given by (10) is valid at all times, but that equation (13) is not valid at the times considered. The behaviour of $S_\alpha(t)$ is consistent with a $1/\sqrt{t}$ convergence to 1 for the system where $F_e$ = 0.5: this is the same rate at which the standard deviation of the distribution of the value of $\bar{\alpha}_t$ goes to zero, however it can also be fitted to other functional forms. The data for the system where $F_e$ =0.01 do not appear to converge to a slope of 1 although it has clearly not reached its limiting behaviour at the times considered.

In Figure 3, we test the FT for *steady state* simulations where neither equation (10) nor (13) is true instantaneously, but we expect (10) to be true for times greater than a several Maxwell times ($\tau_M$=0.084 for T=1.0, n=0.8, $\gamma$=0.5). The steady state trajectory segment averages are obtained from a single trajectory, sampled at different points along the trajectory and consider the same state points as examined for the transient simulations. Clearly, convergence to the limiting behaviour predicted by equation (10) is observed. In contrast, the limiting behaviour predicted by equation (13) has not been realised even at the longest times considered, and the data indicate that *if* convergence does occur, it is very slow compared to its convergence in an isoenergetic system (note that the n = 0.8 state point is quite similar to that studied by Evans, Cohen and Morriss[1] where convergence to within the error bars of the data (3%) was observed when the averaging time was 0.5 and $\gamma$ = 0.5).

The values of $S_\alpha(t)$ determined in the transient experiments are also shown in Figure 3 to demonstrate the convergence of the distribution of the averages over transient trajectories to those over steady state trajectory segments with time. Although (13) is not valid at the times considered, the values of $S_\alpha(t)$ obtained from the transient experiments agree with those obtained from the steady state trajectory segments at sufficiently long times. Furthermore the $S_J(t)$ for the transient and steady state segments agree and equation (10) is verified in both cases. These observations support the possibility that the steady state relationships are correctly predicted from the limiting behaviour of the TFT expressions. It implies that not only does the form of the TFT change when isokinetic dynamics replaces isoenergetic dynamics, but that a corresponding new SSFT is obtained which is given by equation (12).

At equilibrium, equation (13) is clearly false for isokinetic dynamics for any finite t. In this case the probability of observing positive and negative fluctuations in $\overline{\Lambda}_t$ must be *equal* and therefore $S_\alpha(t) = 0$ for all finite times. Equation (13) incorrectly suggests that the logarithm of the probability ratio is proportional to the fluctuation, and that $S_\alpha(t) = 1$. The falsity of equation (13) at equilibrium is consistent with results in figure 3 where the calculated value of $S_\alpha(t)$ is less than unity for the times considered and the departure is greater for the system which is subject to a smaller field. It seems to indicate that either the convergence time becomes infinite as the field goes to zero or that $S_\alpha(t)$ converges to a value less than unity. The behaviour of the fluctuations in $\overline{J}_t$ is correctly predicted by equation (10) which becomes trivial. The observation that equation (13) is not valid for an isokinetic system at equilibrium is not in conflict with Gallavotti and Cohen[4] because the equilibrium system is both non dissipative and non isoenergetic[9] as is required.

We note that it is difficult to test the FT in nonequilibrium systems for long averaging times, large fields or large systems since in these limits, the variance of the distributions approach zero. Therefore, the fluctuations become so narrow that by the time the formula has converged to its asymptotic behaviour, it is not possible to observe the Second Law violating trajectory segments either in a computer simulation or experimentally. This highlights the utility of a Transient rather than the Steady State FT. Tests of the asymptotic Steady State FT can of course *only* be carried out for long averaging times. The Transient FT on the other hand, can be tested for arbitrarily short averaging times where the probability of spontaneous Second Law violations is much greater, approaching 0.5 as $t \to 0$.

The integrated form of the fluctuation formula (IFT) also provides a means of reducing the statistical error when testing the FT for long periods of time.[8] The IFT corresponding to equation (11) is:

$$\frac{p(\bar{J}_t < 0)}{p(\bar{J}_t > 0)} = \left\langle \exp(\bar{J}_t t F_e \beta V) \right\rangle_{\bar{J}_t > 0} = \left\langle \exp(\bar{J}_t t F_e \beta V) \right\rangle_{\bar{J}_t < 0}^{-1} \qquad (16)$$

where $p(\bar{J}_t > 0)$ and $p(\bar{J}_t < 0)$ refer to the probabilities of observing trajectory segments with positive or negative values of $\bar{J}_t$, respectively and the notation $\langle ... \rangle_{\bar{J}_t > 0}$ and $\langle ... \rangle_{\bar{J}_t < 0}$ refer to ensemble averages over trajectories with positive or negative values of $\bar{J}_t$, respectively. We define $Y_J^{'}(t) \equiv (\ln(p(\bar{J}_t > 0) / p(\bar{J}_t < 0)) - 1) / (\ln(\left\langle \exp(\bar{J}_t t F_e \beta V) \right\rangle_{\bar{J}_t < 0}) - 1)$ and in Figure 4 it is verified that $Y_J^{'}(t) = 1$ at all times, in agreement with (16). (Note that subtraction of 1 in the numerator and denominator is carried out to circumvent large statistical errors when the probabilities are almost equal. This property therefore differs from Y(t) used to measure convergence of the IFT in reference [5]). The time that can be considered in the IFT is still limited, particularly at high fields.

By defining $Y_\alpha^{'}(t) \equiv (\ln(p(\bar{\alpha}_t < 0) / p(\bar{\alpha}_t > 0)) - 1) / (\ln(\left\langle \exp(-2N\bar{\alpha}_t t) \right\rangle_{\bar{\alpha}_t > 0}) - 1)$ an IFT corresponding to (13) is

$$\lim_{t \to \infty} Y_\alpha^{'}(t) = \lim_{t \to \infty} \Big( (\ln(p(\bar{\alpha}_t < 0) / p(\bar{\alpha}_t > 0)) - 1) / (\ln \left\langle \exp(-2N\bar{\alpha}_t t) \right\rangle_{\bar{\alpha}_t > 0} - 1) \Big) = 1 \qquad (17)$$

The behaviour of $Y_\alpha^{'}(t)$ for the nonequilibrium systems examined above is shown in Figure 4. Convergence to the limiting behaviour indicated by (17) has not occurred at the longest times considered. An IFT, valid at all times and for arbitrary field strengths is:

$$\frac{p(\bar{\alpha}_t > 0)}{p(\bar{\alpha}_t < 0)} = \left\langle \exp(\bar{J}_t t F_e \beta V) \right\rangle_{\bar{\alpha}_t > 0}^{-1} \qquad (18)$$

This relation predicts that $\tilde{Y}_\alpha^{'}(t) \equiv (\ln(p(\bar{\alpha}_t < 0) / p(\bar{\alpha}_t > 0)) - 1) / (\ln(\left\langle \exp(\bar{J}_t t F_e \beta V) \right\rangle_{\bar{\alpha}_t > 0}) - 1)$ is unity at all times. Note that in a numerical test of equation (18) the large contribution of rarely observed events (when $\bar{J}_t$ is highly positive) to the ensemble average is problematic.[10] This is observed in the data presented in figure 4.

## 4. CONCLUSION

In this paper we have shown how the Transient FT, which was originally derived for the isoenergetic nonequilibrium response of an equilibrium microcanonical ensemble may be generalised to cases where the response is observed for different ensembles with different types of thermostats. The form of the FT may change with ensemble, as shown in Table 1. This should come as no surprise since even at equilibrium fluctuation formulae are generally ensemble dependent.

We have presented numerical results for the response of the isokinetic ensemble to isokinetic shear flow for a system far from equilibrium in the nonlinear regime and for a system in the linear regime. We find that the Transient FT derived here and given by equation (10) for this ensemble/dynamics combination is satisfied for all averaging times. From this Transient FT we obtain a corresponding Steady State FT given by equation (12) for this system. It is observed that the form of the Steady State FT derived from the Transient FT also varies with ensemble. In the limit of long averaging times the computational results again support the validity of the proposed Steady State FT for this combination of ensemble and dynamics.

## ACKNOWLEDGEMENTS

We would like to thank the Australian Research Council for the support of this project. Helpful discussions and comments from Professor E.G.D. Cohen and Dr. Gary Ayton are also gratefully acknowledged.

## References.

**Figure Captions**

**Figure 1**. The normalised logarithmic probability ratio W(X) for the fluctuations in a transient response numerical experiment with T = 1.0, n = 0.8, N=32, $\gamma$ = 0.5 and a transient trajectory segment of length t = 0.6. The circles are a test of the transient FT given by equation (11) and the crosses show that the limiting expression given by equation (13) is not applicable, at least at this value of t. The straight line is the expected result from equation (11) and (13).

**Figure 2**. The squares show the slope of straight line fitted through a plot of $W_J(t)$ versus $\bar{J}_t$, as a function of t for the transient response of a 2 dimensional system of 32 particles at a) T = 1.0, n = 0.8 and $\gamma$ = 0.5 and b) T = 1.0, n = 0.4 and $\gamma$ = 0.01. Equation (11) predicts a slope of 1 at all times, which is indeed observed. The circles show the slope of a straight line fitted through a plot of $W_\alpha(t)$ versus of $\bar{\alpha}_t$, as a function of t for the same system. Equation (13) says the slope will be 1 in the limit of long times. O(1/N) effects have been accounted for.

**Figure 3**. The squares show the slope of straight line fitted through a plot of $W_J(t)$ versus $\bar{J}_t$, as a function of t for a steady state simulation of a 2 dimensional system of 32 particles at a) T = 1.0, n = 0.8 and $\gamma$ = 0.5 and b) T = 1.0, n = 0.4 and $\gamma$ = 0.01. Equation (11) predicts a slope of 1, which is consistent with the results. The circles show the slope of a straight line fitted through a plot of $W_\alpha(t)$ versus of $\bar{\alpha}_t$, as a function of t for the same system. The crosses show how the slope of a straight line fitted through a plot of $W_\alpha(t)$ versus $\bar{\alpha}_t$ varies with t using data for a transient experiment (as also shown in figure 2). Convergence of the transient and steady state results is observed. O(1/N) effects have been accounted for.

**Figure 4**. Tests of the integrated fluctuation formulae given by equations (16)-(18) for a 2 dimensional system of 32 particles at a) T = 1.0, n = 0.8 and $\gamma$ = 0.5 and b) T = 1.0, n = 0.4 and $\gamma$ = 0.01. The squares show $Y_J^{'}(t)$ which is predicted to be unity at all times by equation (16). If equation (17) is correct, then $Y_\alpha^{'}(t)$, shown by the circles, should approach unity at long times. The crosses show $\tilde{Y}_\alpha^{'}(t)$, which will equal unity at all times if (18) is valid. Large numerical errors

in $\tilde{Y}'_{\alpha}(t)$ are observed for the system at large field, as expected.[10]

**Table I.** Transient fluctuation formula in various ergodically consistent ensembles.[a,b]

| | | |
|---|---|---|
| Isokinetic dynamics | $\ln\frac{\Pr(\bar{J}_t)}{\Pr(-\bar{J}_t)} = -\bar{J}_t t F_e \beta V$ | $-JF_eV = \frac{dH_0^{ad}}{dt}$ |
| Isobaric-isothermal | $\ln\frac{\Pr(\bar{J}_t)}{\Pr(-\bar{J}_t)} = -\bar{J}_t t F_e \beta V$ | $-JF_eV = \frac{dI_0^{ad}}{dt}$ |
| Isoenergetic | $\ln\frac{\Pr(\overline{J\beta}_t)}{\Pr(-\overline{J\beta}_t)} = -\overline{J\beta}_t t F_e V$ or $\ln\frac{\Pr(\bar{\Lambda}_t)}{\Pr(-\bar{\Lambda}_t)} = -\bar{\Lambda}_t t$ | $-JF_eV = \frac{dH_0^{ad}}{dt}$ |
| Isoenergetic boundary driven flow | $\ln\frac{\Pr(\bar{\Lambda}_t)}{\Pr(-\bar{\Lambda}_t)} = -\bar{\Lambda}_t t$ | |
| Nosé-Hoover (canonical) dynamics | $\ln\frac{\Pr(\bar{J}_t)}{\Pr(-\bar{J}_t)} = -\bar{J}_t t F_e \beta V$ | $-JF_eV = \frac{dH_0^{ad}}{dt}$ |
| Wall ergostatted field driven flow[c] | $\ln\frac{\Pr(\overline{J\beta_{wall}}_t)}{\Pr(-\overline{J\beta_{wall}}_t)} = -\overline{J\beta_{wall}}_t t F_e V$ or $\ln\frac{\Pr(\bar{\Lambda}_t)}{\Pr(-\bar{\Lambda}_t)} = -\bar{\Lambda}_t t$ | $-JF_eV = \frac{dH_0^{ad}}{dt}$ |
| Wall thermostatted field driven flow[c] | $\ln\frac{\Pr(\bar{J}_t)}{\Pr(-\bar{J}_t)} = -\bar{J}_t t F_e \beta V - \ln\left(\left\langle \exp\left[\bar{\Lambda}_t t(1-\beta_{system}/\beta_{wall})\right]\right\rangle_{\bar{J}_t}\right)$ | $-JF_eV = \frac{dH_0^{ad}}{dt}$ |
| Steady state isoenergetic dynamics:[d] $\bar{J}_t = \frac{1}{t}\int_{t_0}^{t_0+t} J(s)ds$ where $t_0 >> \tau_M$ | $\ln\frac{\Pr(\overline{J\beta}_t)}{\Pr(-\overline{J\beta}_t)}$ $= -\overline{J\beta}_t t F_e V - \ln\left(\left\langle \exp\left[F_e V(\int_0^{t_0} J(s)\beta(s)ds + \int_{t_0+t}^{2t_0+t} J(s)\beta(s)ds)\right]\right\rangle_{\bar{J}_t}\right)$ $\lim_{t\to\infty}\frac{1}{t}\ln\frac{\Pr(\overline{J\beta}_t)}{\Pr(-\overline{J\beta}_t)} = -\overline{J\beta}_t F_e V$ | $-JF_eV = \frac{dH_0^{ad}}{dt}$ |

[a] It is assumed that the limit of large system has been taken so that O(1/N) effects can be neglected. In the Nosé-Hoover dynamics, it is also assumed that the mass associated with the heat bath is an extensive variable.

[b] $H_0$ is the equilibrium internal energy and $I_0$ is the equilibrium enthalpy.

[c] In the wall ergostatted/thermostatted systems, it is assumed that the energy/temperature of the full system (wall and fluid) is fixed.

[d] Similar steady state formula can be obtained for other ensembles. $\tau_M$ is the Maxwell time that characterises the time required for relaxation of the nonequilibrium system into a steady state.

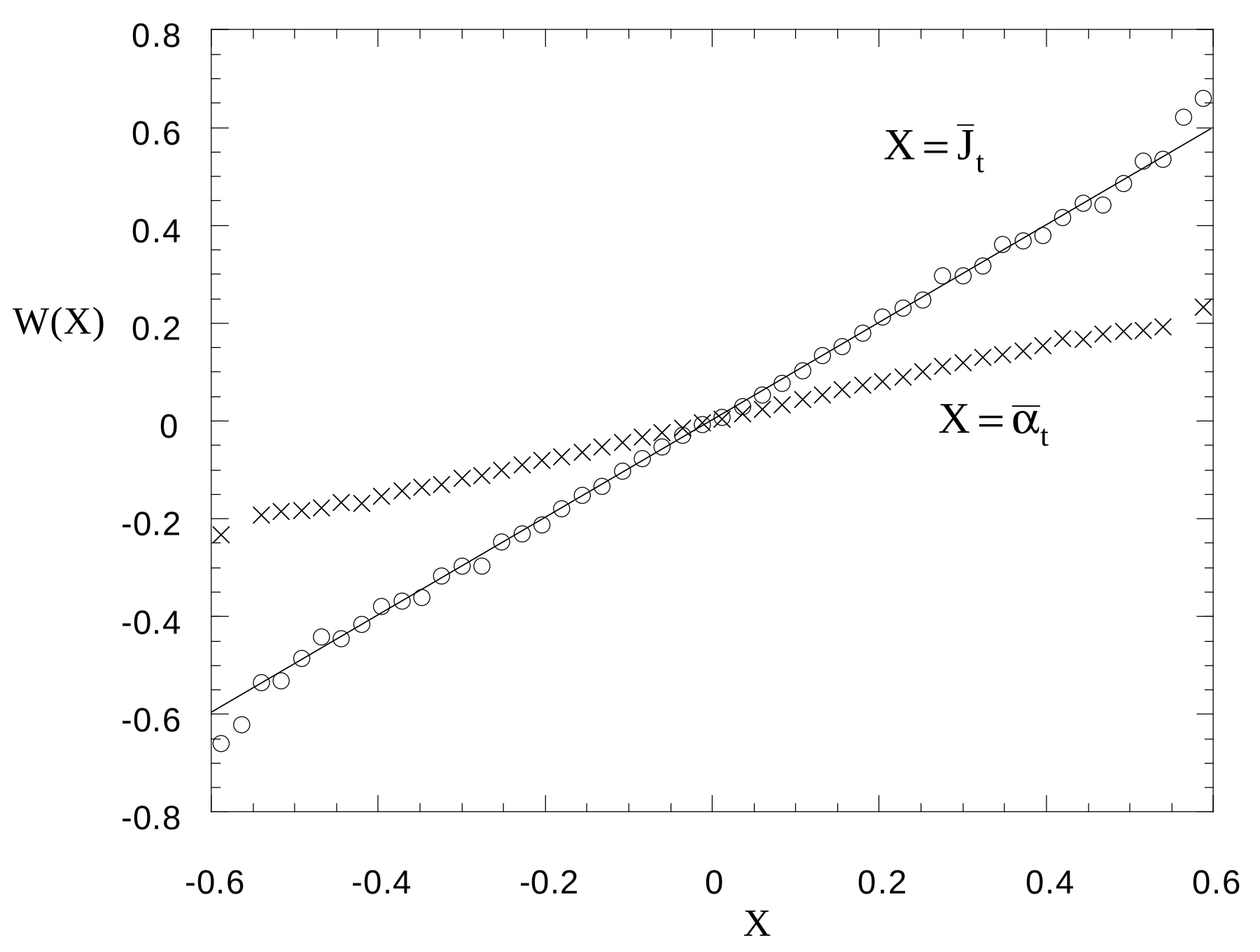

Figure 1

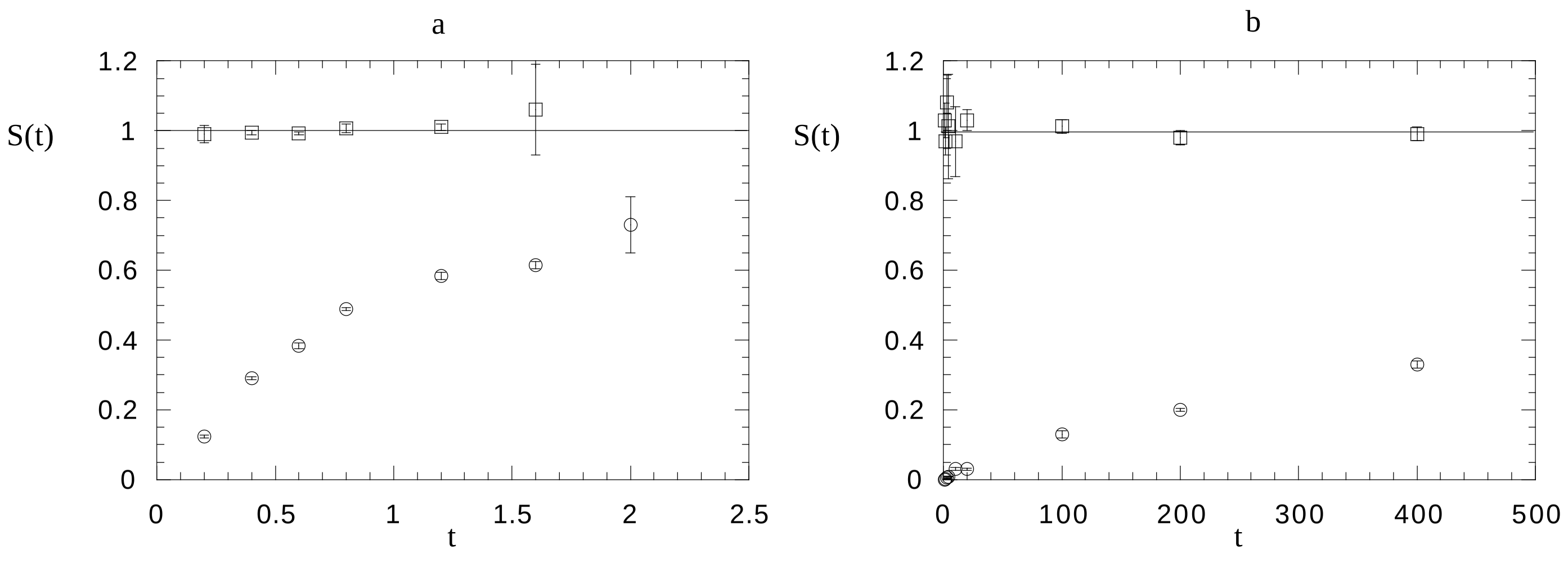

Figure 2

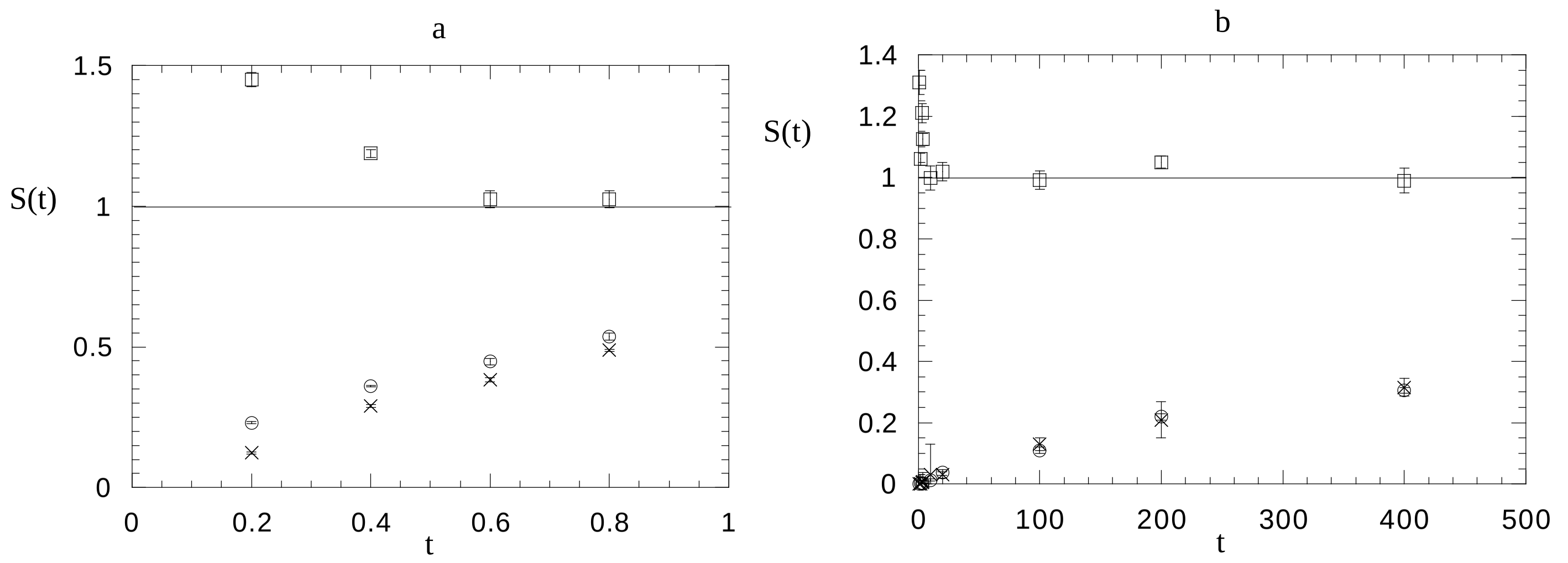

Figure 3

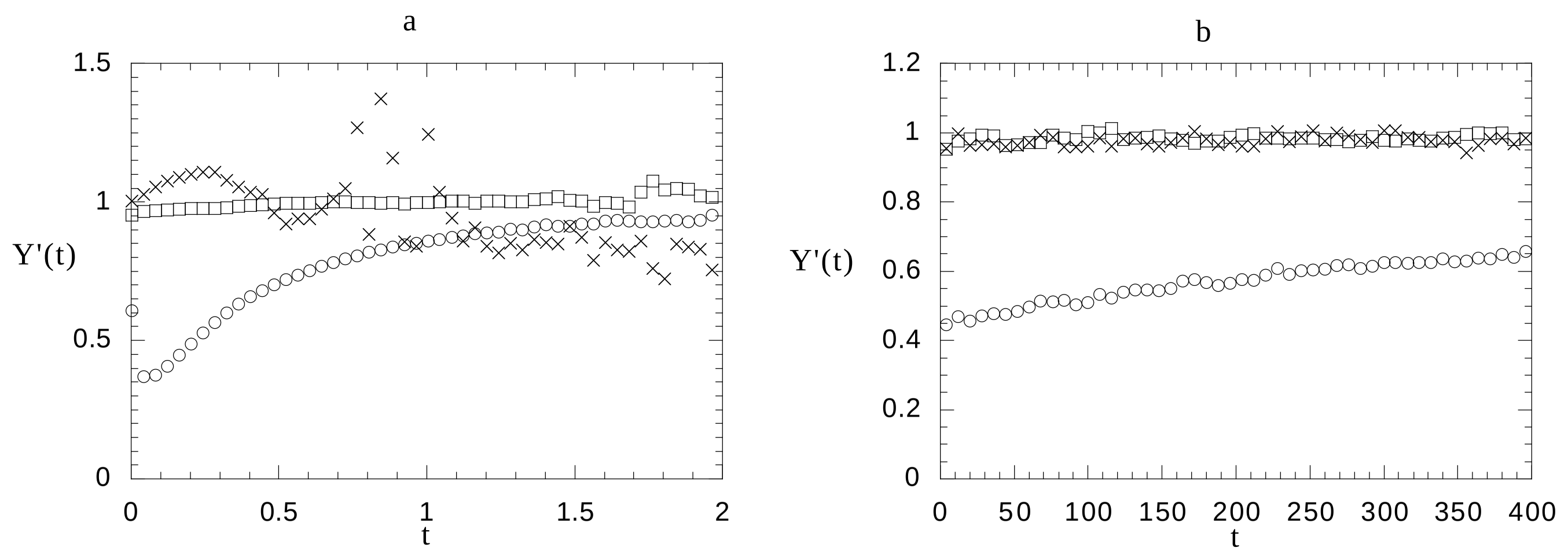

Figure 4